# METHODS OF MEASURING PHYSICAL CHARACTERISTICS OF PARTIAL COMPONENTS OF MULTICOMPONENT SAMPLES

**A.Yu. Buki**

*National Science Center "Kharkov Institute of Physics & Technology",*
*1, Akademicheskaya St., Kharkov 61108, Ukraine*
*E-mail: abuki@ukr.net*

Consideration is given to the methods of gaining experimental data on the substances which constitute a part of multicomponent samples to be measured. The methods are applicable to the samples comprising an arbitrary number of components; their use is not restricted to single-type experiments only. The application of the methods is demonstrated with an example of obtaining the spectra of electrons scattered by $^6$Li and $^7$Li nuclei, which were measured from two targets having different isotopic compositions. As a result of the proposed methods, spectra were obtained for the electrons scattered by the nuclei of each of the constituent isotopes, as if the experiment were made on isotopically pure targets.

In a number of experiments, one of the essential requirements on the object to be studied is the level of its chemical or isotopic purity required for the given study. It is not uncommon that the purity of obtainable substances limits the possibility of their investigation to the required accuracy. The reasons for the nonavailability of the necessary purity of materials may be different. Among them are: a high cost of the materials, a technical unattainability of the required purification level or, for example, the nuclear-safety restrictions in case of work with highly-enriched isotopes of fissionable materials, etc. Of course, the technological progress makes the high-purity materials more attainable, yet at the same time the demands on the accuracy of experimental results also increase. Therefore, the conflict between the experimental requirements and the availability of required purity materials is everlasting in character.

We cite a few examples of using impure materials in electronuclear experiments. Thus, in some measurements on titanium and iron nuclei, targets of natural isotopic composition have been used (see p. 126 in the monograph [1]). As a result, the data obtained may be considered only conditionally as the ones related to the $^{48}$Ti and $^{56}$Fe isotopes that make up 43.99% and 91.66% in the target, respectively. A similar situation is observed in the measurements on lithium isotopes, where the isotopic composition of targets was 90.5% $^6$Li in one case [2], and 93.8% $^7$Li in the other case [3]. In the studies of $^{112}$Sn [4], the target material was enriched with this isotope up to 70%, and in the case of $^{114}$Sn – up to 61%. It is obvious that handling the problem in the experiments which use impure materials, lies in correct taking into account the contribution from impurities to the data measured. However, in the present paper, this problem is formulated in a more general form.

The aim of the work is to consider the problem of calculating the contributions to the data measured, which come from all the components comprised in the sample under study.

### *Analysis and measurements*

Let us consider the relationship between the measured data and the composition of the sample under study. Assume that $i$ is the number of one of the $I$ samples under study; $j$ is the number of one of $J$ components constituting the sample; $\delta_{i,j}$ is the portion of the component $j$ in the sample $I$; $A_i^{exp}$ is the result of measurement of the parameter $A$ on the sample $i$.

All values of $\delta_{i,j}$ and $A_i^{exp}$ are known. Unknown are the characteristics $A_j$ of each of the components $j$ making up the sample. Unless the characteristic $A_j$ of each of the components is dependent on the presence of other components in the sample, the measurement result can be written as

$$A_i^{\exp} = \sum_{j=1}^{J} A_j \delta_{i,j} \ . \qquad (1)$$

For $I$ samples we write down the following set of equations

$$\begin{cases} A_1\delta_{1,1} + ... + A_j\delta_{1,j} + ... + A_J\delta_{1,J} = A_1^{\exp} \\ ...................................................... \\ A_1\delta_{i,1} + ... + A_j\delta_{i,j} + ... + A_J\delta_{i,J} = A_i^{\exp} \\ ...................................................... \\ A_1\delta_{I,1} + ... + A_j\delta_{I,j} + ... + A_J\delta_{I,J} = A_I^{\exp} \end{cases} . \quad (2)$$

The solution of this set of equations relative to the $A_j$ values can be found with the help of the known computer mathematics programs. With the aim of determining the conditions, at which the solution of the system is unique, it is necessary to consider the matrix of the following form:

$$A = \begin{pmatrix} \delta_{1,1} & ... & \delta_{1,j} & ... & \delta_{1,J} \\ & & ............................... & & \\ \delta_{i,1} & ... & \delta_{i,j} & ... & \delta_{i,J} \\ & & ............................... & & \\ \delta_{I,1} & ... & \delta_{I,j} & ... & \delta_{I,J} \end{pmatrix} . \quad (3)$$

The uniqueness of solution of the system of equations (2) requires that in the matrix (3) there should be $I = J$ and the determinant $|A| \neq 0$.

These requirements lead to the following restrictions on the conditions of sample measurements:

1) the number of samples $I$ measured should be equal to the number of components $J$ making up these samples;
2) the component proportions should be different in all the samples.

As soon as the $A_j$ values are determined, there arises the question about their errors, i.e., $\Delta A_j$. Generally, the errors in determining the quantitative composition of the material, i.e., the $\delta_{i,j}$ values, are considerably smaller than the measurement errors of $A_i^{exp}$, i.e., $\Delta A_i^{exp}$. Therefore, with a statistical character of $\Delta A_i^{exp}$, the relation of this error to the unknown errors $\Delta A_j$ is given by

$$\left(\Delta A_i^{\exp}\right)^2 = \sum_{j=1}^{J} \left(\Delta A_j\right)^2 \left(\delta_{i,j}\right)^2 . \quad (4)$$

Equations (4) and (1) differ only in the exponent of quantities entering into them. Thus, the solution of the set of equations of form (4) with respect to the unknown $(\Delta A_j)^2$ is found in the same way as the solution to the set of equations (2) with respect to $A_j$.

We shall call the proposed approach to the measurement of multicomponent samples as the technique of component characteristics separation (CCS).

## The occurrence of molecular substance in the samples

We consider the case, where the samples comprise, aside from the components, the quantity of which is arbitrary, some not a single-element molecular substance (hereafter referred to as "substance X"), which consists of $n$ chemical elements. So, the proportion of different elements in the substance X is specified by its chemical formula, i.e., it is invariable, and the variations in the quantity of the substance X itself in the samples do not change the proportions of the X elements in each of the samples. Mathematically, this means that in the determinant $A$ there will appear $n$ similar columns, and therefore, the determinant will be equal to zero, and the corresponding set of equations will not have the unique solution. In other words, in the case considered, the CCS technique is inapplicable and it appears impossible to determine the characteristics of all the components entering into the substance X. This raises the question as to whether it is possible to measure the characteristics of other components of the samples, which do not enter into the substance X.

In all rows of determinant (3) the proportion between the substance X elements is the same. That is, in the $i \neq 1$ sample the portion of all substance X elements differs by a factor of $k_i$ from the portions of these elements in the $i=1$ sample, or $\delta_{i,p} = k_i\delta_{1,p}$ for each $p$, where $p$ denotes the indices of elements entering into the substance X. The $k_i$ value may vary from sample to sample. Hence, if for the $i=1$ sample the contribution to the sum $A_1^{exp}$ (eq. (1)) of the substance X is denoted by $X_1$, then this contribution to the sum $A_i^{exp}$ of arbitrary sample can be written as

$$X_i = k_i X_1 , \quad (5)$$

and eq. (1) can be presented in the form

$$A_i^{\exp} = k_i X_1 + \sum_{j \neq p} A_j \delta_{i,j} \,. \qquad (6)$$

Since we assume that all $\delta_{i,j}$ values are known, then $k_i = \delta_{i,p}/\delta_{1,p}$ and eq. (6) leads to the system of equations with $J - n + 1$ unknowns $X_1$ and $A_j$ ($j \neq p$).

It is interesting to note the case, where certain elements enter into the substance X and, at the same time, are the separate components of the samples. For example, if we take $CuSO_4$ as a substance X, and the samples contain the metal Cu as a separate component, then in eq. (6) each of them will have its own characteristics $X_1$ and $A_j$ and the system of equations will have the solution as in the previous case.

**Application of the proposed technique in nuclear physics experiments**

The application of the CCS methods holds promise for nuclear physics experiments since they are concerned with the measurements of nuclear reaction cross sections, which are independent of the cross sections of other nuclei present in the target. The reaction cross sections present the nuclear characteristic. Therefore, the cross-section determination includes the normalization to the number of nuclei in the target, and in the case of a not single-component target, the cross section $\sigma_i^{exp}$, measured on it, will be the weighted sum of cross sections $\sigma_j$ for all the nuclides comprised in the target:

$$\sigma_i^{\exp} = \sum_j \delta_{i,j} \sigma_j \,, \qquad (7)$$

where $\delta_{i,j}$ is the portion of the nuclide $j$ by the number of nuclei in the target. This equation is the same as eq. (1), and therefore, if measurements of $\sigma_i^{exp}$ are made on several targets having different component ratios, then the measured data can be analyzed using the CCS technique.

In nuclear physics experiments, it is rather common to meet the problem of obtaining targets of high isotopic purity. The process of enrichment of natural substance with one of its constituent isotopes is rather complicated and labor-consuming. Besides, the cost of enriched substance sharply increases with a reduction in the content of other isotopes. The CCS technique makes it possible to perform the experiment on relatively low-enriched targets, but it calls for the measurements on several targets. Therefore, the choice between the measurements on one highly enriched target or the measurements on several less enriched targets is determined by the ratio of the cost of an extra highly-enriched material to the cost of additional time of experimental setup operation. We note two cases when this dilemma is solved in favor of the CCS application: a) the measurements at moderate-size particle accelerators, where the operating time cost is relatively low; b) if the aim of the investigation is not one but a few isotopes making up the target.

We give the example of using the CCS technique in the experiment on electron scattering by the $^6$Li and $^7$Li nuclei. In the measurements, use was made of the two targets having the following weight contents of isotopes: $^6$Li $\delta'_{1,1}$=0.905, $^7$Li $\delta'_{1,2}$= 0.095 (target № 1) and $^6$Li $\delta'_{2,1}$=0.062, $^7$Li $\delta'_{2,2}$= 0.938 (target № 2). In the experiment, cross-sections for electron scattering by the nuclei were measured. Therefore, as mentioned above, it is necessary to use the portion of the number of nuclei δ rather than the weight fraction of the isotope in the target δ'. It is not difficult to derive the equation relating these parameters for the target № i, which contains two types of nuclides 1 and 2

$$\delta_{i,1} = \left[ 1 + \frac{\delta'_{i,2}}{\delta'_{i,1}} \frac{M_1}{M_2} \right]^{-1} , \qquad (8)$$

where $M$ is the atomic mass. In the case of two nuclides we have $\delta_{i,2} = 1 - \delta_{i,1}$. Using eq. (8), we obtain the $\delta_{i,j}$ values of the system of eqs. (2), where $A_i^{exp}$ are the cross sections $\sigma_i^{exp}$ measured on the targets $i$ = 1, 2, and $\sigma_1$ and $\sigma_2$ are the unknown cross-sections on the nuclei $^6$Li and $^7$Li, respectively. The solution of this set of two equations has the form

$$\sigma_1 = \frac{\sigma_1^{\exp}\delta_{2,2} - \sigma_2^{\exp}\delta_{1,2}}{\delta_{1,1}\delta_{2,2} + \delta_{1,2}\delta_{2,1}}$$
$$\sigma_2 = \frac{\sigma_2^{\exp}\delta_{1,1} - \sigma_1^{\exp}\delta_{2,1}}{\delta_{1,1}\delta_{2,2} + \delta_{1,2}\delta_{2,1}} \qquad (9)$$

Figure 1 shows the spectra of electrons scattered by the lithium isotope nuclei of target № 1 (upper plot) and target № 2 (bottom plot). After applying eq. (9) to each pair of points measured on targets № 1 and № 2 at the same energy $E'$, two spectra were obtained, one of which corresponds to the measurements on the target with a 100% content of $^6$Li, and the other spectrum – to the measurements on the 100% $^7$Li target.

In the given spectra, measured with a low energy resolution, the effect of CCS application is most conspicuous in the region of the elastic electron scattering peak (Fig. 1, peak el). The energy position of the elastic scattering peak depends on the nuclear mass. In the spectra discussed for the case of $^6$Li and $^7$Li nuclei, the energies of the peak differ by 1 MeV (see Fig. 1). At measurements on the isotopically impure target, the contribution from the impurity isotope nuclei shifts the experimental peak towards the peak corresponding to scattering by the impurity nuclei only.

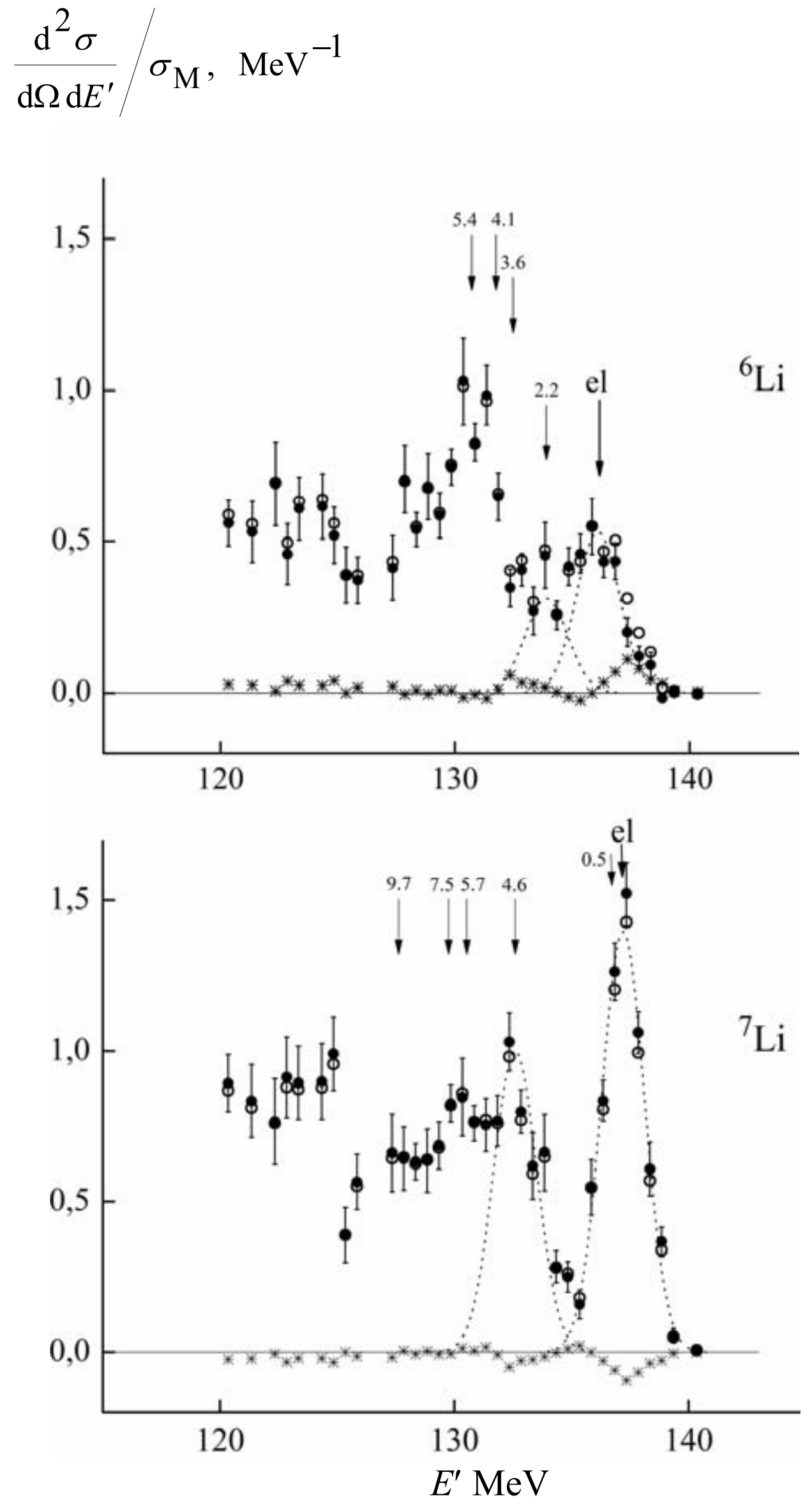

*Fig. 1. Spectra of differential cross-sections of electrons scattered by the $^6$Li and $^7$Li nuclei, normalized to the cross-section for scattering by the unit point charge, $\sigma_M$. The initial electron energy is 144 MeV, the scattering angle is 160°. The open circles show the results of measurements, the closed circles - the same data after their correction by eqs. (9), the asterisks show the difference between the initial and corrected data. The dotted curves are drawn around the elastic scattering peak (el) and the peak of the first excited state, the contribution of which is noticeable in the spectrum.*

It follows that the application of eqs. (9) to the spectrum points, which were measured under the same conditions on two targets having different isotopic compositions, must result in the shift of elastic scattering peaks to the opposite directions. Precisely this effect can be seen in Fig. 1 if you compare the points on the right-hand side of elastic scattering peaks before and after their correction by means of eqs. (9). Apart from this effect, the correction discussed may result in essential corrections to the areas under the peaks, from which the cross section for the elastic electron scattering by nuclei is found. Thus, for the area under the peak of elastic scattering by $^6$Li nuclei (see Fig. 1) the correction makes 11%, while the statistical measurement error of this area is 5%. In case of $^7$Li nuclei, the correction in question makes 5%, while the statistical error is 3.5%.

We now consider the possibility of taking into account the effect of not a single-element molecular impurity present in the targets on the results of the measurements. In the lithium case, this may be the corrosion-preventive compound that comes from the manufacturer. The targets are made by press forming from supplied unformed pieces of metal. The process of press forming takes place in air medium, and therefore, because of a very quick lithium oxidability, the corrosion-preventive compound cannot be fully removed, and some amount of it remains pressed inside the targets. After the measurements have been made, the amount of that compound remaining in each of the targets can be exactly determined by chemical analysis. Since the chemical formula of the compound specifies the proportion of elements entering into its composition and the targets comprise one and the same slushing compound, this proportion is the same in all the targets. So, we come to the case of eq. (6), when the coefficients $k_i$ in this equation are found, as stated above, by means of the chemical analysis, and $X_{I,P}$ is one of the unknowns. In the case under consideration, apart from the measurements on targets № 1 and № 2, the solution of the equation calls for the measurements on a certain target № 3. The isotopic composition of this target is of no importance here, but the content of the corrosion-preventive compound in it must be different from that in targets № 1 and № 2. This can be attained by consciously introducing a greater amount of the compound into the target № 3 during its press forming than the amount comprised in targets № 1 or № 2.

## Results and conclusions

Based on the measurements of multicomponent samples, the technique of component characteristics separation has been proposed for determining the characteristics of the sample constituent components. Or, in other words, the technique has been offered, which allows one to determine the contributions to the measured data from all the components comprised in the sample under study.

The proposed technique is applicable in the case, where the investigated characteristic of each of the components is independent of the presence of other components in the sample (e.g., mechanical mixture of components).

In nuclear physics investigations on the isotopes, two cases can be distinguished, when the present technique is obviously efficient:

a) in the measurements at the setups, the operating time cost of which is not high in comparison with the cost of the material having ultrahigh enrichment of the isotope under study;

b) if the aim of the investigation is more than one isotope present in the material.